\documentclass{PoS}

\title{Studies of the nature of the low-energy, gamma-like background for Cherenkov Telescope Array}

\ShortTitle{Low-energy, gamma-like background of CTA}
\author{\speaker{Julian Sitarek}\\
        University of \L\'od\'z, PL-90236 Lodz, Poland\\
        E-mail: \email{jsitarek@uni.lodz.pl}}
\author{Dorota Sobczy\'nska\\
        University of \L\'od\'z, PL-90236 Lodz, Poland\\
        E-mail: \email{dsobczynska@uni.lodz.pl}}
\author{Micha\l{} Szanecki\\
        University of \L\'od\'z, PL-90236 Lodz, Poland\\
        E-mail: \email{mitsza@uni.lodz.pl}}
\author{Katarzyna Adamczyk\\
        University of \L\'od\'z, PL-90236 Lodz, Poland\\
        E-mail: \email{kadamczyk@uni.lodz.pl}}
\author{for the CTA Consortium}

\abstract{
The upcoming Cherenkov Telescope Array (CTA) project is expected to provide unprecedented sensitivity in the low-energy ($\lesssim100$\,GeV) range for Cherenkov telescopes. 
In order to exploit fully the potential of the telescopes the standard analysis methods for gamma/hadron separation might need to be revised. 
We study the composition of the background by identifying events  composed mostly of a single electromagnetic subcascade or double subcascade from a $\pi^0$ (or another neutral meson) decay. 
We apply the standard simulation and analysis chain of CTA to evaluate the potential of the standard analysis to reject such events.
}

\FullConference{35th International Cosmic Ray Conference --- ICRC2017-\\
		10--20 July, 2017\\
		Bexco, Busan, Korea}

\def\ses{\emph{SES}}
\def\spis{\mbox{\emph{S}$\pi^0$\emph{S}}}
\newcommand{\progname}[1]{{\fontfamily{pcr}\selectfont #1}}

\begin{document}

\section{Introduction}
The imaging air Cherenkov technique has been successfully used for observations of a $\gamma$-ray emission from astrophysical sources since nearly 30 years. 
The principle of the technique is based on the measurement of Cherenkov photons produced in the atmosphere by the charged relativistic particles from an Extensive Air Shower (EAS). 
The two dimensional angular distribution of Cherenkov light forms the shower image on the telescope camera. 
Even for bright sources, the number of registered hadron-induced events (the so-called background) is a few orders of magnitude larger than the number of registered $\gamma$-rays events.
Therefore the $\gamma$/hadron separation method plays the crucial role in the analysis of data from Cherenkov telescopes. 
An effective $\gamma$-ray selection is obtained using the Hillas image parameterization \cite{hi85}. 
More sophisticated selection methods (see e.g. \cite{kraw2006,al08,ohm09,par14}) are being used nowadays, but most of them are still based on the original Hillas parameters.

Over the years, the construction of larger mirror dish telescopes and the employment of the stereoscopic technique allowed to lower the observation energy threshold. 
Currently three large Imagining Air Cherenkov Telescope (IACT) instruments are in operation: H.E.S.S. \cite{aha06}, MAGIC \cite{al16a} and VERITAS \cite{hold11}.
The upcoming Cherenkov Telescope Array (CTA) \cite {actis11,acha13} was designed to study $\gamma$-ray sources in a broad energy range, from a few tens of~GeV to hundreds of~TeV. 
CTA is expected to bring an  order of magnitude improvement in the sensitivity with respect to the current IACT systems \cite{be13}. 
Nevertheless, at low energies the $\gamma$/hadron separation becomes more difficult, which results in the deterioration of the sensitivity. 
Such deterioration is a combined effect of instrumental effects and the underlying physics of the EAS. 
The former is mostly connected with smaller and dimmer images of low-energy showers.
As the shower has to be reconstructed from the information in only a few pixels, the reconstruction performance is degraded. 
On the other hand several physical processes may affect the observations at the lowest energies as well. 
As such energies the geomagnetic field has more impact on $\gamma$ rays (making them appear more hadron-like) than hadron initiated showers thus the efficiency of primary  particle selection is worse (see e.g. \cite{bo92,sz13}). 
Second, at low energies larger fluctuations of the image parameters are expected due to the larger fluctuations of the Cherenkov light density at the ground \cite{bhat}. 
Third, a primary electron or positron can induce an EAS that can mimic a $\gamma$-ray shower (see e.g. \cite{co01}). 
Finally, $\gamma$-ray events may be imitated by a specific type of a hadron-induced shower. 
It has been suggested in \cite{maier} that hadronic events that survive the $\gamma$-ray selection criteria, have transferred a large fraction of the primary's energy to electromagnetic sub-cascades during the first few interactions in EAS. 
Furthermore, a large telescope can be triggered by light solely produced by $e^\pm$ from only a single or double electromagnetic sub-cascades, produced in a decay of a $\pi^0$  \cite{sob2007,so15a}.
Such events (caused mainly by protons with $E\lesssim 200$\,GeV) will produce images with very similar shapes to $\gamma$-ray events, contributing to the $\gamma$-like background. 

In this contribution we study with Monte Carlo (MC) simulations the impact of such events on the observations with the CTA-North array. 
In particular, we test the ability of the state-of-the-art CTA analysis methods to reject such a background. 

\section{Simulations and analysis}
We simulated $\gamma$-ray and proton showers using a code of \progname{CORSIKA} 7.5 \cite{he98}, that was adapted for the purpose of our study. 
The geophysical parameters of the chosen La Palma site were selected.
The telescope array consists of 4 LSTs (Large Size Telescopes) and 15 MSTs (Medium Size Telescopes) following one of the promising CTA-North layouts, \progname{3AL4M15-5}. 
The response of the telescopes was then simulated using the \progname{sim\_telarray} code \cite{be08} with the individual telescope parameters following the so-called \progname{Production-3} settings.
The results of the simulations were processed with \progname{Chimp} \cite{ha15,ha17} to allow analysis using MAGIC Analysis and Reconstruction Software (MARS) \cite{za13,al16b}.

\begin{figure}[t!]
\centering
\begin{tabular}{cc}
\begin{minipage}[b]{0.49\textwidth}
\centering
    \includegraphics[width=0.95\textwidth]{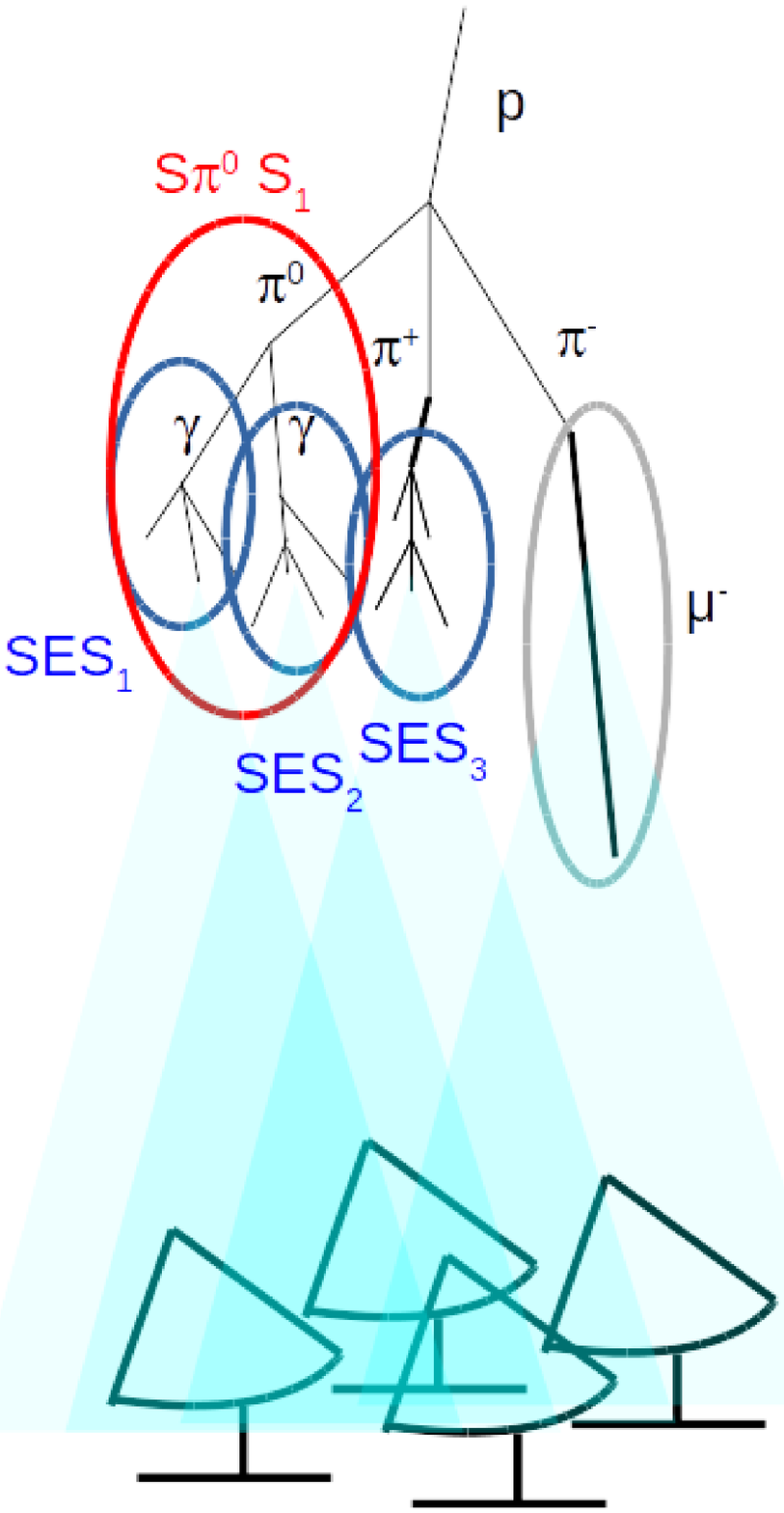}
\caption{
Schematic representation of an example hadronic shower in which $\pi^\pm$ and $\pi^0$ are produced. 
$\pi^0$ decays into two $\gamma$ rays initiating two separate \ses\ and one \spis .
$\pi^+$ decays into $\mu^+$, which in turn decays into $e^+$ initiating third \ses .
$\pi^-$ decays into $\mu^-$ which in this case continues through the atmosphere. 
\vspace{0.95cm}}
\label{fig:shower}
\end{minipage}
&
\begin{minipage}[b]{0.49\textwidth}
\centering
    \includegraphics[width=0.8\textwidth]{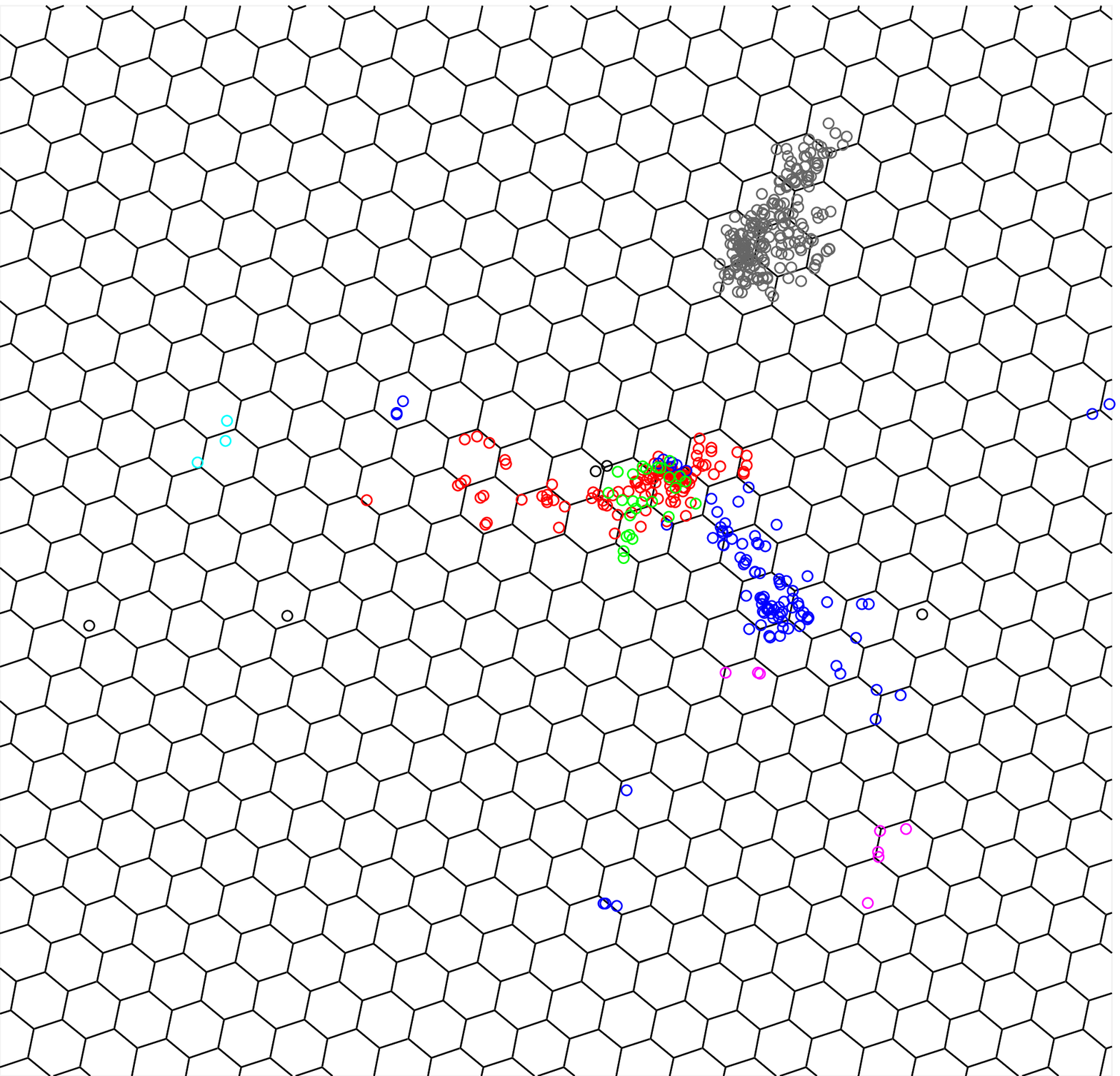}\\ \vspace{0.5cm}
    \includegraphics[width=0.8\textwidth]{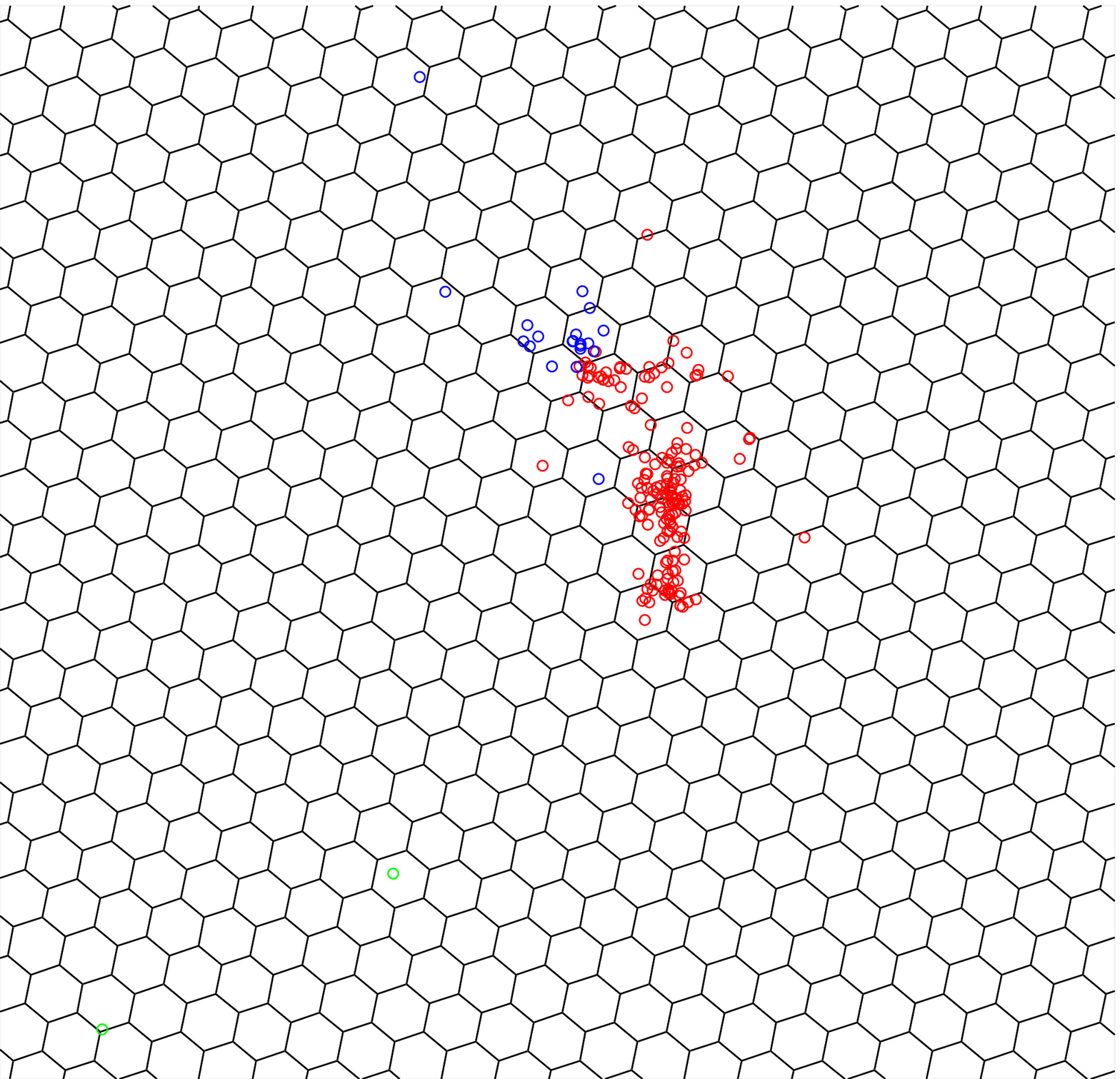}
    \caption{Example image of an event with multiple \ses\ (top) and a \ses - dominated event (bottom). 
Each circle represents the position of a single phe in a LST camera coordinates. 
Different colours represent different \ses . 
Gray points show phe produced by non-\ses\ component (in this case by muons).
Only a part of the camera containing the shower image is shown.
Black hexagons show individual pixels of the camera. }
    \label{fig:image}
\end{minipage}
\end{tabular}
\end{figure}

During the simulations of EAS we mark the occurrence of a Single Electromagnetic Subcascade, hereafter \ses\ and a Single $\pi^0$ Subcascade, hereafter \spis\ , in the shower (see Fig.~\ref{fig:shower}).
We define \ses\ as a particle (normally e$^\pm$ or a $\gamma$-ray) taking part in an electromagnetic interaction, and all the secondary particles created in the electromagnetic cascade starting from that particle. 
Similarly, we define \spis\ as the primary particles created in a decay of a neutral particle (typically $\pi^0$ or $\eta$) and all the secondary particles created in the subshower started by these particles. 
We assign a unique number to each new \ses\ and \spis\ produced in the shower. 
Then, each Cherenkov photon induced by a $e^\pm$ has two additional numbers propagated, to identify the \ses\ and \spis\ to which it belongs.

For the Cherenkov photons that reach the telescope camera and are converted into photoelectrons (phe) we calculate the statistics of \ses\ and \spis .
A \ses\  (or \spis ) is considered to contribute to the event if it produced at least 6\,phe in at least one of the triggered telescopes. 
For each \ses\ that satisfy the above condition we compute a ratio of a number of phe originating from it to the total number of phe measured in all the triggered telescopes. 
We call \ses$_{\max}$ the largest of these ratios (i.e. for the most dominating \ses ), and similarly \spis$_{\max}$ for the most dominating \spis .
An event is \ses -dominated if \ses$_{\max}>70\%$ and  \spis -dominated if \spis$_{\max}>70\%$. 
This information is then propagated throughout the analysis chain. 
In Fig.~\ref{fig:image} we show an example image composed of multiple \ses\ and a \ses -dominated event.

%
Different \ses\ can form different parts of the image (compare e.g. red and blue points in the top panel of Fig.~\ref{fig:image}), resulting in irregular, easy to reject events. 
Parts of the image produced by different \ses\ might be also registered at similar angular direction (compare green and red points in the same panel).
In case of events with one dominating \ses\ (see the bottom panel of Fig.~\ref{fig:image}), the image is more regular and thus will be able to imitate $\gamma$-rays more effectively.

We perform the $\gamma$/hadron separation and energy estimation using multidimensional decision trees, the so-called Random Forest (RF) method \cite{al08}. 
RF is using Hillas parameters of a given image (size, width, length, fraction of size in two brightest pixels) together with stereo reconstruction parameters from the whole event (impact and the height of the shower maximum) as well as estimated energy of the event.

To evaluate the effect of \ses\ and \spis\ on typical observations we define G80 cuts, i.e. a cut in \emph{Hadronness} that at a given estimated energy preserves 80\% of $\gamma$ rays. 
In order to investigate background for typical CTA sources and to avoid camera edge effects we apply a cut in the reconstructed source position. 
Only the background events with reconstructed position withing 1.5$^\circ$ from the camera center are considered in the analysis. 
In such region the angular acceptance is nearly constant. 
Proton events are reweighted to their cosmic ray spectra, i.e a power-law with a spectral slope of $-2.73$. 
$\gamma$ rays are reweighted to a power-law with a spectral slope of $-2.6$. 

\section{Results}
\begin{figure*}[t]
\centering
\includegraphics[width=0.49\textwidth]{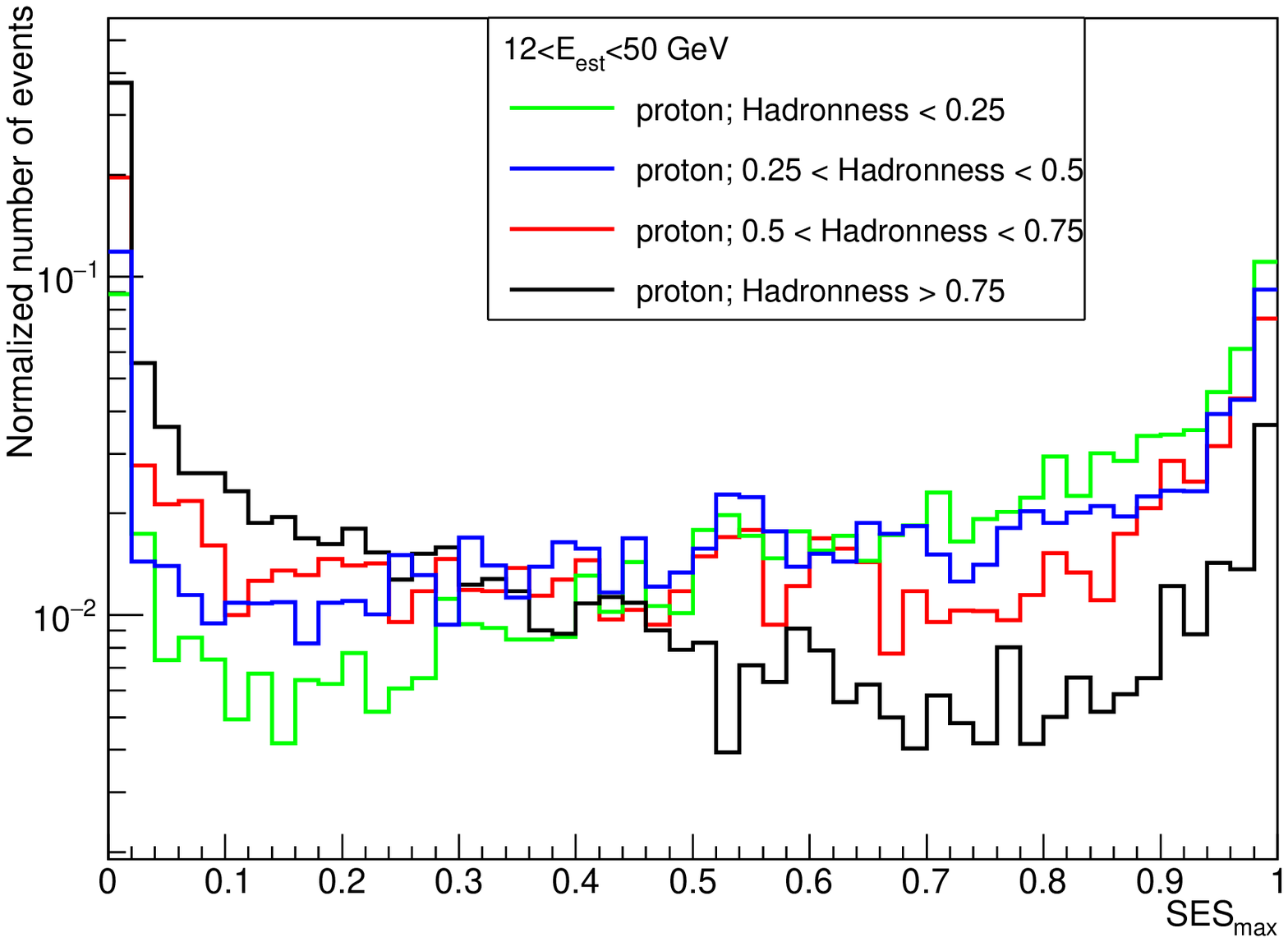}
\includegraphics[width=0.49\textwidth]{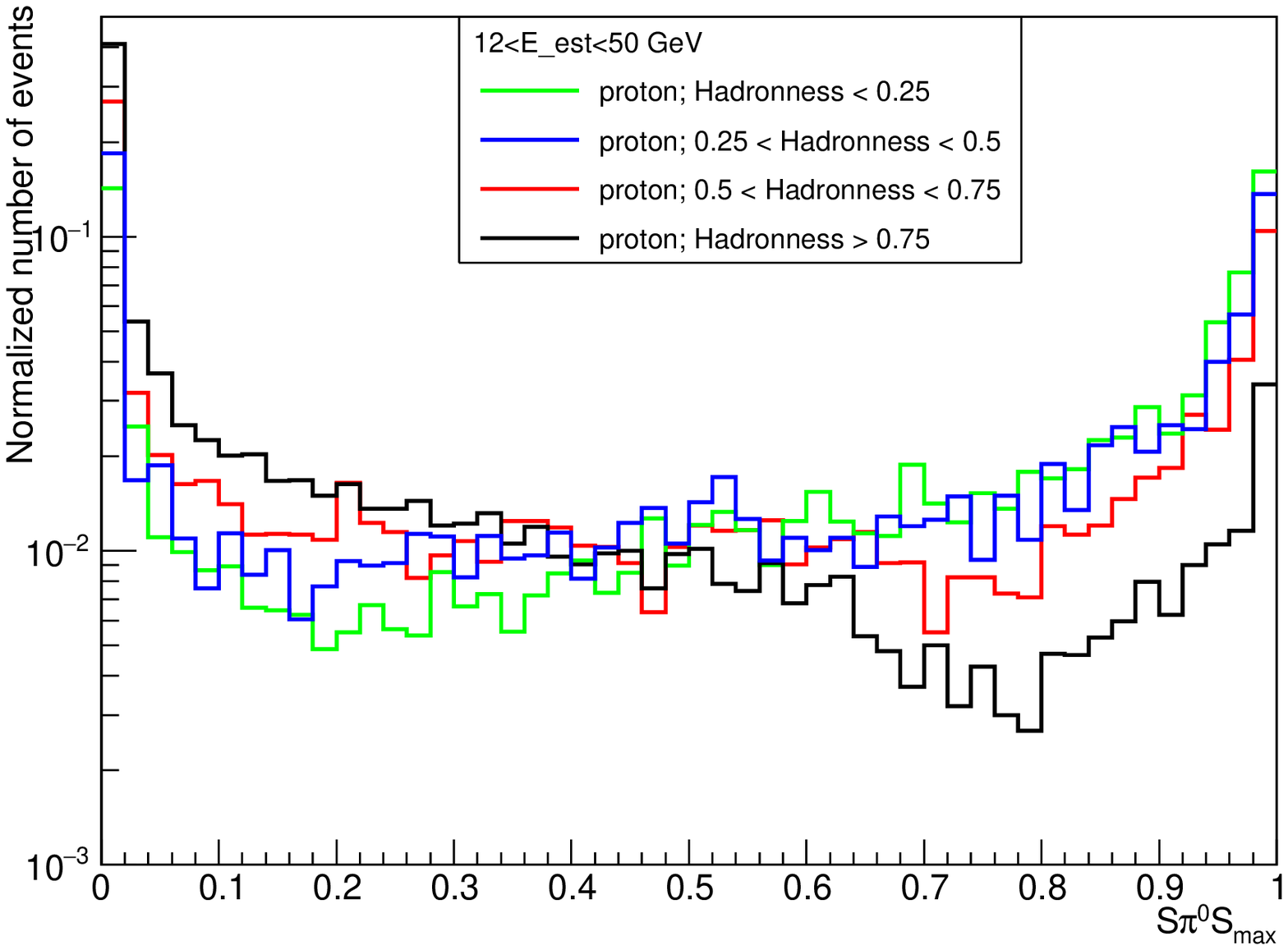}
\caption{Distribution of \ses$_{\max}$ (the left panel) and \spis$_{\max}$ (the right panel) for proton events in different bins of \emph{Hadronness} (see legend). 
LST sub-array is used and only events with estimated energy between 12 and 50 GeV are plotted. 
}\label{fig:vssmax}
\end{figure*}

In Fig.~\ref{fig:vssmax} we show the distributions of \ses$_{\max}$ and \spis$_{\max}$ parameters for different bins of the aggregated $\gamma$/hadron separation parameter, \emph{Hadronness} for the lowest energies accessible to the LST sub-array. 
Events with low \emph{Hadronness} value have often high \ses$_{\max}$ (and \spis$_{\max}$). 
Comparing the two panels of Fig.~\ref{fig:vssmax}, the \spis$_{\max}\approx1$ peak for low \emph{Hadronness} values is more pronounced than the corresponding peak at \ses$_{\max}\approx1$. 
The events that enhance this peak in the former case are most probably composed of a single \spis\ containing two \ses\ of comparable size.
Peak at \ses$_{\max}$=\spis$_{\max}$=0 is composed mostly of muon-dominated events. 

\begin{figure*}[t]
\centering
\includegraphics[width=0.49\textwidth]{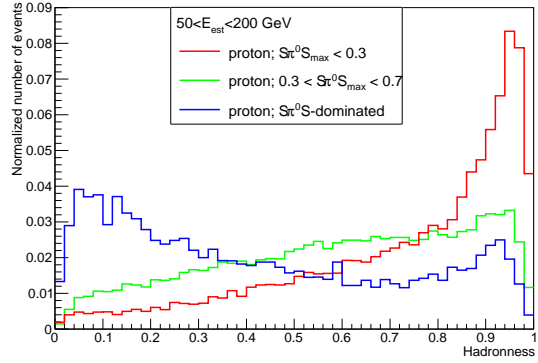}
\caption{Distribution of \emph{Hadronness} for \spis -dominated (\spis$_{\max}>0.7$, blue) proton events compared with proton events without a dominating \spis\ ($0.3<$\spis$_{\max}<0.7$, green and \spis$_{\max}<0.3$, red). 
The estimated energy range of 50-200\,GeV for the full (MST+LST) array was used.
}\label{fig:vshadr}
\end{figure*}

We compute the distribution of \emph{Hadronness} for events with a different dominance of the largest \spis\ (see Fig.~\ref{fig:vshadr}).
A clear difference in \emph{Hadronness} distribution between \spis -dominated and \spis -not-dominated events is visible. 
The former produce a peak at low values and thus efficiently imitate showers initiated by $\gamma$ rays. 
On the other hand, events without a dominating \spis\ are classified with a \emph{Hadronness} value mainly close to 1 and thus are easily rejected from the analysis.

\begin{figure}[t]
\includegraphics[width=0.49\textwidth]{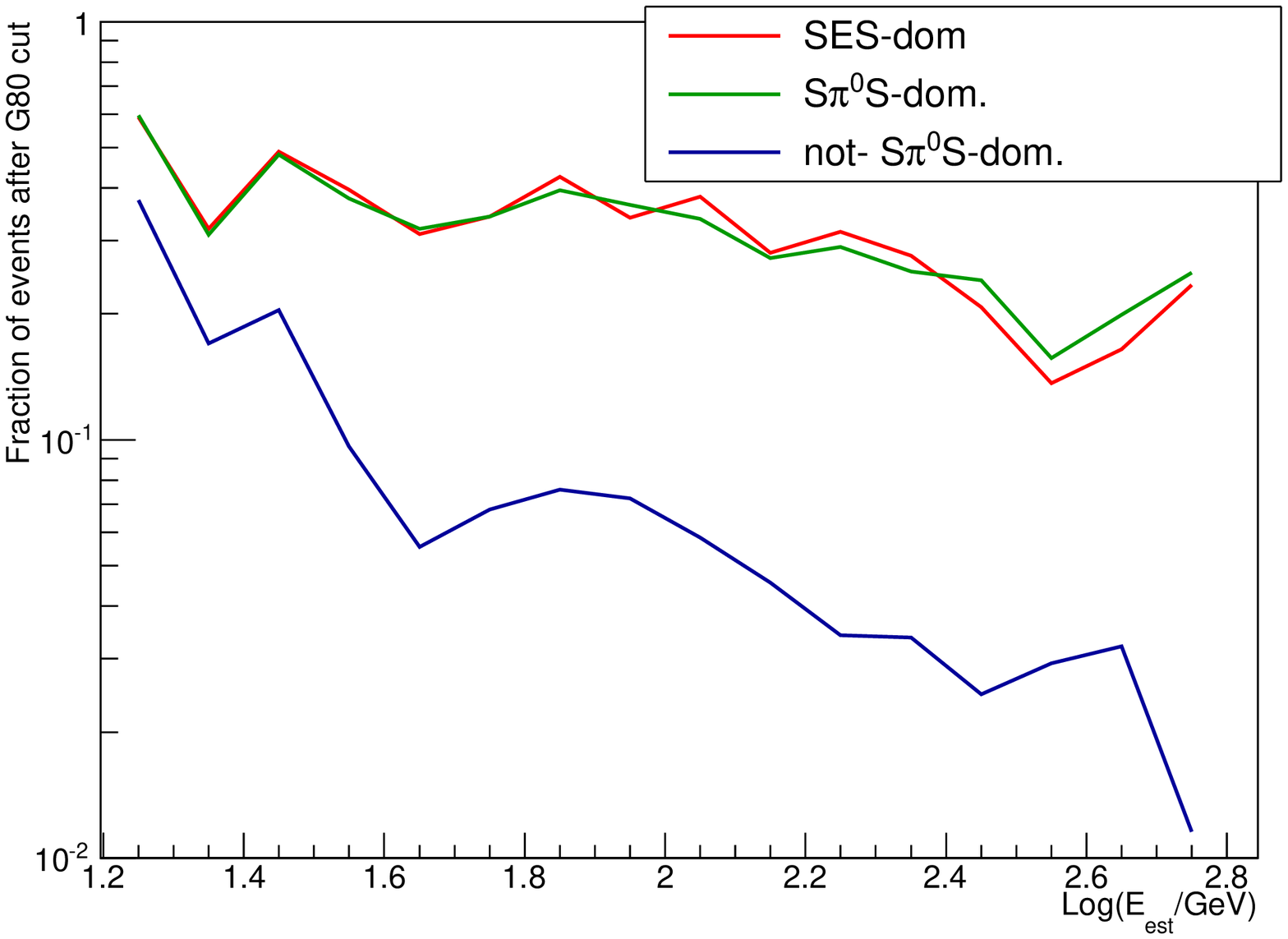}
\includegraphics[width=0.49\textwidth]{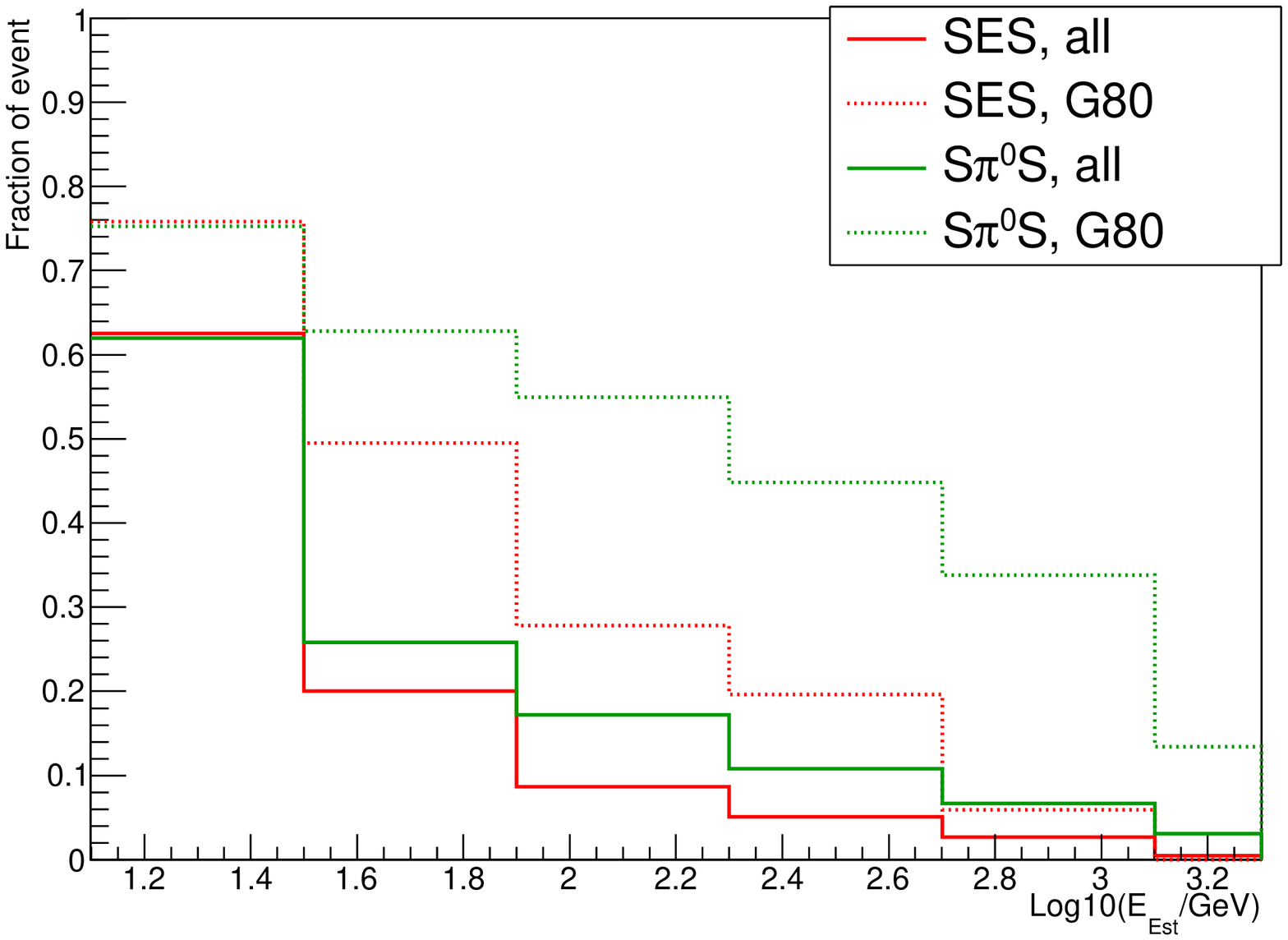}
\caption{
Left panel: Energy dependent fraction of events surviving G80 cut among following groups: \ses -dominated (red), \spis -dominated (green), and non \spis -dominated (blue)
Right panel: energy dependent fraction of \ses -dominated (red) and \spis -dominated (green) events before a \emph{Hadronness} cut (solid), and after G80 cut (dotted). 
Full array is used. 
}\label{fig:frac}
\end{figure}

In the left panel of Fig.~\ref{fig:frac} we present the separation power for different classes of events. 
Both the \ses -dominated and \spis -dominated events are difficult to distinguish from $\gamma$-ray initiated shower. 
Only about 60\% of such events are rejected with a G80 cuts (note that those cuts reject also 20\% of $\gamma$ rays). 
This is nearly an order of magnitude worse than for events without a dominating \spis\ and improves only very slowly with energy.

In the right panel of Fig.~\ref{fig:frac} we present how the fraction of \ses -dominated  and \spis -dominated events changes with the estimated energy.
The fraction of both \ses - and \spis -dominated events decreases fast with increasing energy. 
This can be explained as a larger number of individual \ses\ and \spis\ are produced in a higher energy shower and can be observed by the telescopes. 
As the showers composed of multiple \ses\ or \spis\ are much easier to separate, after a cut in \emph{Hadronness} the fraction of \ses - and \spis -dominated events is much higher.
At 100\,GeV it reaches 34\% and 57\% respectively.  
It is interesting to note that, despite about twice larger fraction of \spis -dominated events than \ses -dominated events, the separation power of both types of events is very similar. 
This suggests that such single-\spis -double-\ses\ events are still similar to a single \ses -dominated events and thus hard to separate from primary $\gamma$ rays. 

\section{Conclusions}
Using standard CTA simulation software and state-of-the-art Cherenkov telescopes analysis methods we have studied low energy background events of CTA. 
In particular, we have investigated events composed mainly of a single electromagnetic subcascade, or a pair of electromagnetic subcascades produced in a decay of a neutral particle. 
We performed full MC simulations of $\gamma$ rays and protons for one of the most promising arrays designed for CTA-North. 
As expected \ses -dominated and \spis -dominated showers are very similar to $\gamma$-ray induced showers, and hence difficult to reject.
After $\gamma$/hadron separation cuts the influence of the \ses - and \spis -dominated events relatively increases.
They constitute $\gtrsim$50\% of the residual cosmic ray background, mimicking $\gamma$ rays with energies $\lesssim 100$\,GeV.

\section*{Acknowledgements}
\sloppy
This work is supported by the grant through the Polish Narodowe Centrum Nauki No. 2015/19/D/ST9/00616.
DS is supported by the Narodowe Centrum Nauki grant No. 2016/22/M/ST9/00583. 
This work was conducted in the context of the CTA Analysis and Software Working Group.
We gratefully acknowledge financial support from the agencies and organizations listed here: 
\texttt{http://www.cta-observatory.org/consortium\_acknowledgments}
We would like to thank CTA Consortium and MAGIC Collaboration for allowing us to use their software. 
We would also like to thank Abelardo Moralejo, Paolo Cumani and Konrad Bernl\"ohr for helpful discussions.

\end{document}